\documentclass[a4paper,fleqn]{cas-sc}

\usepackage[authoryear,longnamesfirst]{natbib}

\def\tsc#1{\csdef{#1}{\textsc{\lowercase{#1}}\xspace}}
\tsc{WGM}
\tsc{QE}

\begin{document}
\let\WriteBookmarks\relax
\def\floatpagepagefraction{1}
\def\textpagefraction{.001}

\newcommand{\textbit}[1]{\textbf{\textit{#1}}}
\newcommand{\inspace}[1]{\vspace*{#1}\noindent}
\newcommand{\sqrtsNN}{\ensuremath{\sqrt{s_\mathrm{NN}}}}
\newcommand{\pTjet}{\ensuremath{p_\mathrm{T,jet}}}
\newcommand{\pT}{\ensuremath{p_\mathrm{T}}}
\newcommand{\vtwo}{\ensuremath{v_{2}}}
\newcommand{\ICP}{\ensuremath{I_\mathrm{CP}}}
\newcommand{\RAA}{\ensuremath{R_\mathrm{AA}}}
\newcommand{\IAA}{\ensuremath{I_\mathrm{AA}}}
\newcommand{\gammadir}{\ensuremath{\gamma_\mathrm{dir}}}
\newcommand{\pizero}{\ensuremath{\pi^{0}}}
\newcommand{\GeVc}
{GeV/{\it c}}

\shorttitle{Jet quenching overview at ATHIC 2025}    

\shortauthors{N. R. Sahoo}  

\title [mode = title]{Jet Quenching in Heavy-Ion Collisions at RHIC and the LHC experiments}   



%
\author{\textcolor{black}{Nihar Ranjan Sahoo }}[
       orcid=0000-0003-4518-6630,
]




\ead{niharsahoo@iisertirupati.ac.in}



\affiliation{organization={Indian Institute of Science Education and Research (IISER)},
            city={Tirupati},
            postcode={517619}, 
            state={AP},
            country={India}}




\begin{abstract}
Jet quenching serves as a key probes of the Quark-Gluon Plasma (QGP) in heavy-ion collisions. This proceedings presents recent results from RHIC and LHC on jet energy loss, acoplanarity, and the flavour and path-length dependence of Parton energy loss, providing critical constraints on QGP properties and theoretical models. Upcoming data taking campaigns at RHIC and the LHC will offer enhanced precision and extended kinematic reach to further advance our understanding of jet-medium interactions. 
\end{abstract}



\maketitle

\section{Introduction}\label{}
At extreme temperatures and/or nuclear densities, ordinary hadrons transition into a hot and dense medium governed by Quantum Chromodynamics (QCD), known as the quark-gluon plasma (QGP), where color degrees of freedom become deconfined and play a dominant role. Heavy-ion collision experiments at RHIC and the LHC provide a unique opportunity to study the QGP and its interactions with high-energy partons as they traverse the medium. These highly virtual partons, produced in high-$Q^{2}$ processes, fragment into collimated sprays of hadrons known as jets. In heavy-ion collisions, jet quenching—manifested as the suppression of high-momentum hadrons and jets, along with modifications to jet substructure relative to pp collisions—serves as a fundamental signature of QGP formation~\cite{PHENIX:2006avo,STAR:2003pjh,ALICE:2019qyj,ALargeIonColliderExperiment:2021mqf,STAR:2016jdz}. Understanding jet–medium interactions is essential for probing QGP transport properties, such as its opacity and coupling strength.\\

This proceeding discusses four key manifestations of jet–medium interactions observed in heavy-ion collision experiments at RHIC (STAR) and the LHC (ALICE, CMS, ATLAS). Specifically, it examines the dependence of jet yield suppression on the jet resolution parameter ($R$), differences in suppression between quark- and gluon-initiated jets, intra-jet broadening, jet acoplanarity, and modifications to jet substructure. While these phenomena are discussed in detail in the following sections, the topic of jet substructure modification is not covered extensively in this proceeding.\\

Before presenting recent results on jet quenching, Section~\ref{Sec:Challenges} outlines the experimental challenges associated with jet measurements in heavy-ion collisions and briefly summarizes the strategies used to address them. Section~\ref{sec:jetquenchingObservations} highlights key recent observations related to jet quenching, and Section~\ref{Sec::summary} concludes with a summary and outlook.

\begin{figure}
	\begin{center}		
		\includegraphics[width=0.7\columnwidth]{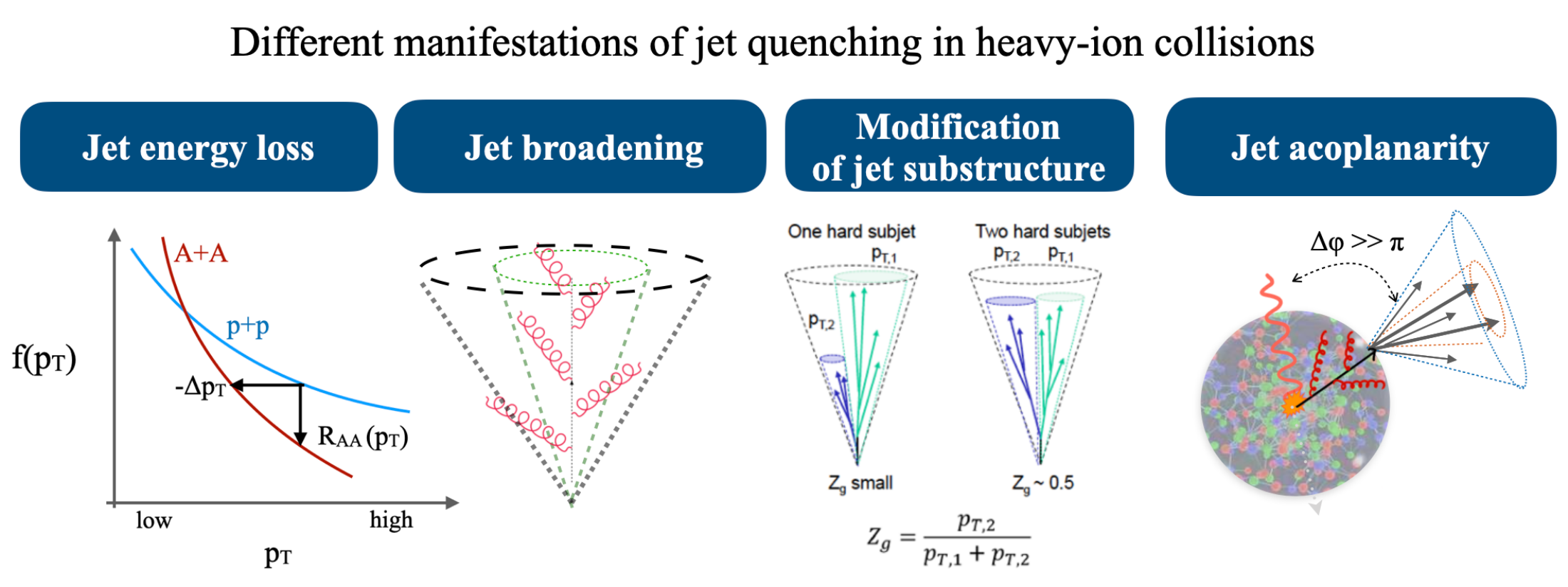}
        
		\caption{Four manifestations of jet quenching in heavy-ion collisions.} 
        \label{fig:jetQManifest}
	\end{center}
\end{figure}

\section{Addressing jet measurement challenges in heavy-ion collisions}\label{Sec:Challenges}

Jet reconstruction in heavy-ion collisions involves using infrared and collinear (IRC) safe algorithms—such as the $k_{T}$, anti-$k_{T}$ ~\cite{Cacciari:2008gp} and Cambridge/Aachen (C/A) ~\cite{Dokshitzer:1997in} algorithm—to reliably identify and cluster particles into jets within a defined cone size, typically set by the radius parameter $R$, and with low transverse momentum (\pT) cutoff. In heavy-ion collisions due to large number of produced tracks from bulk of the medium introduces a significant challenge: uncorrelated background particles (combinatorial jets) originating from the underlying event can obscure the signal from hard-scattered (high-$Q^{2}$) partons. This uncorrelated jet background is particulalry dominated at low jet momentum. To address this, recent measurements have utilized data driven method---the semi-inclusive hadron+jet and \gammadir+jet measurements using the mixed event method in STAR ~\cite{STAR:2017hhs,STAR:2023pal,STAR:2023ksv} and the trigger sample subtraction method ~\cite{ALICE:2015mdb} in ALICE. The latter method effectively removes the multi-parton interactions in the jet sample. These approaches enable the measurment of semi-inclusive recoil jets down to relatively low transverse momentum (\pTjet\ $>\approx$ 5\GeVc) in heavy-ion collisions.\\

 On the other hand, experimental measurements are constrained by detector resolution---such as tracking efficiency, \pT-resolution, Calorimeter performance. This necessitates the use of unfolding techniques and corrections to account for inefficiencies and smearing. Jet observables such as shapes, substructure, and fragmentation patterns are particularly sensitive to these detector effects and uncorrelated background effect, as well as to non-perturbative QCD processes. \\

Over the years, experiments at RHIC and the LHC have developed increasingly sophisticated techniques to address these challenges. This proceeding highlights selected physics results without delving into the details of the analysis methods in the following sections.

\section{Recent jet quenching measurements}
\label{sec:jetquenchingObservations}
A hard-scattered parton traversing the QGP medium interacts with it via soft gluon radiation in addition to its vacuum-like cascade. As a consequence, jet yields are suppressed in heavy-ion collisions relative to those in $pp$ collisions. The additional in-medium radiation leads to intra-jet broadening, which can also manifest as modifications in the jet substructure. Furthermore, interactions with the medium can cause the jet axis to deviate from the original direction of the highly virtual parent parton—a phenomenon known as jet acoplanarity. These manifestations are illustrated in Fig.~\ref{fig:jetQManifest}. The following section discusses each of these effects in detail.
\begin{figure}[htb]
	\begin{center}		
		\includegraphics[width=0.4\columnwidth]{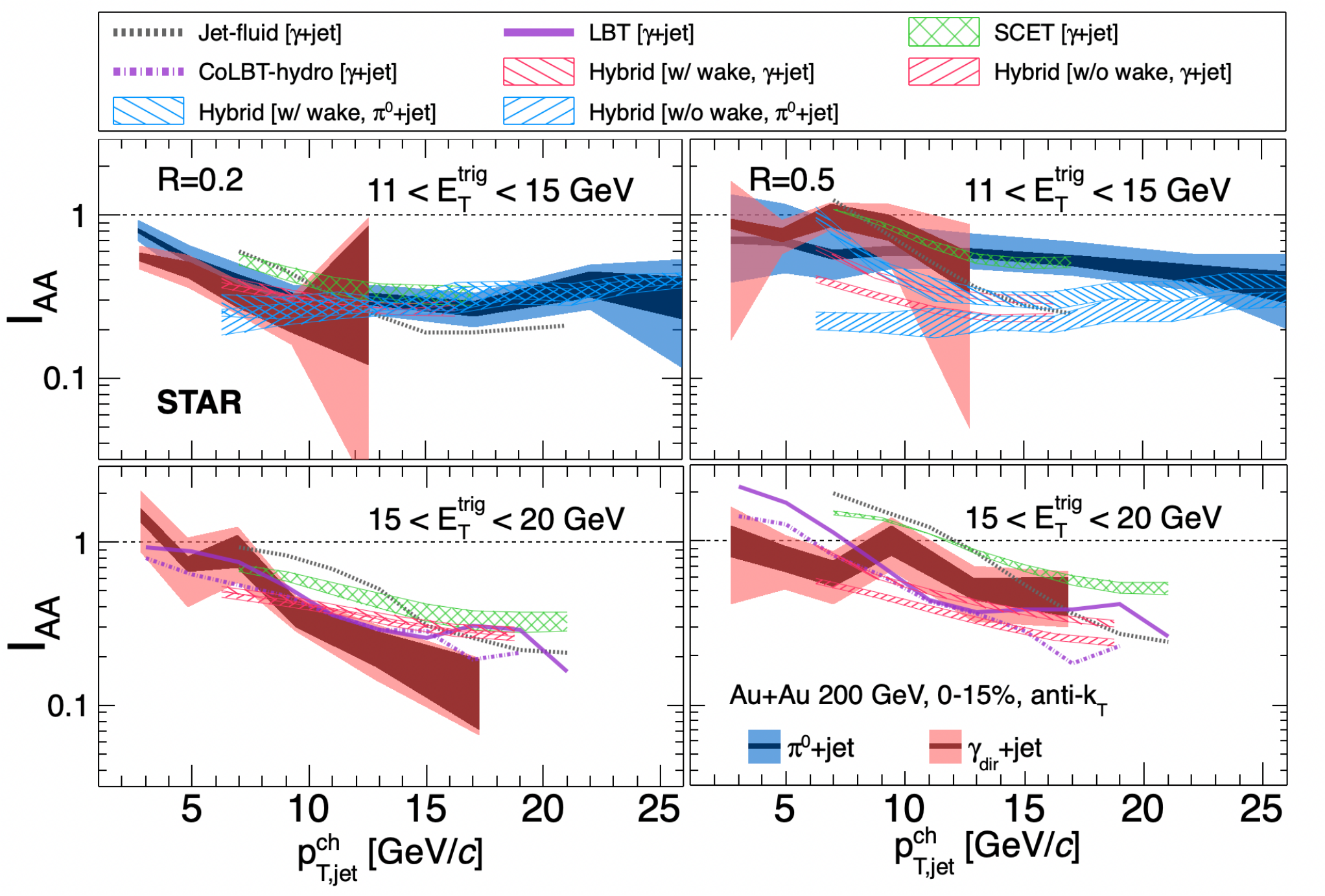}
        \includegraphics[width=0.3\columnwidth]{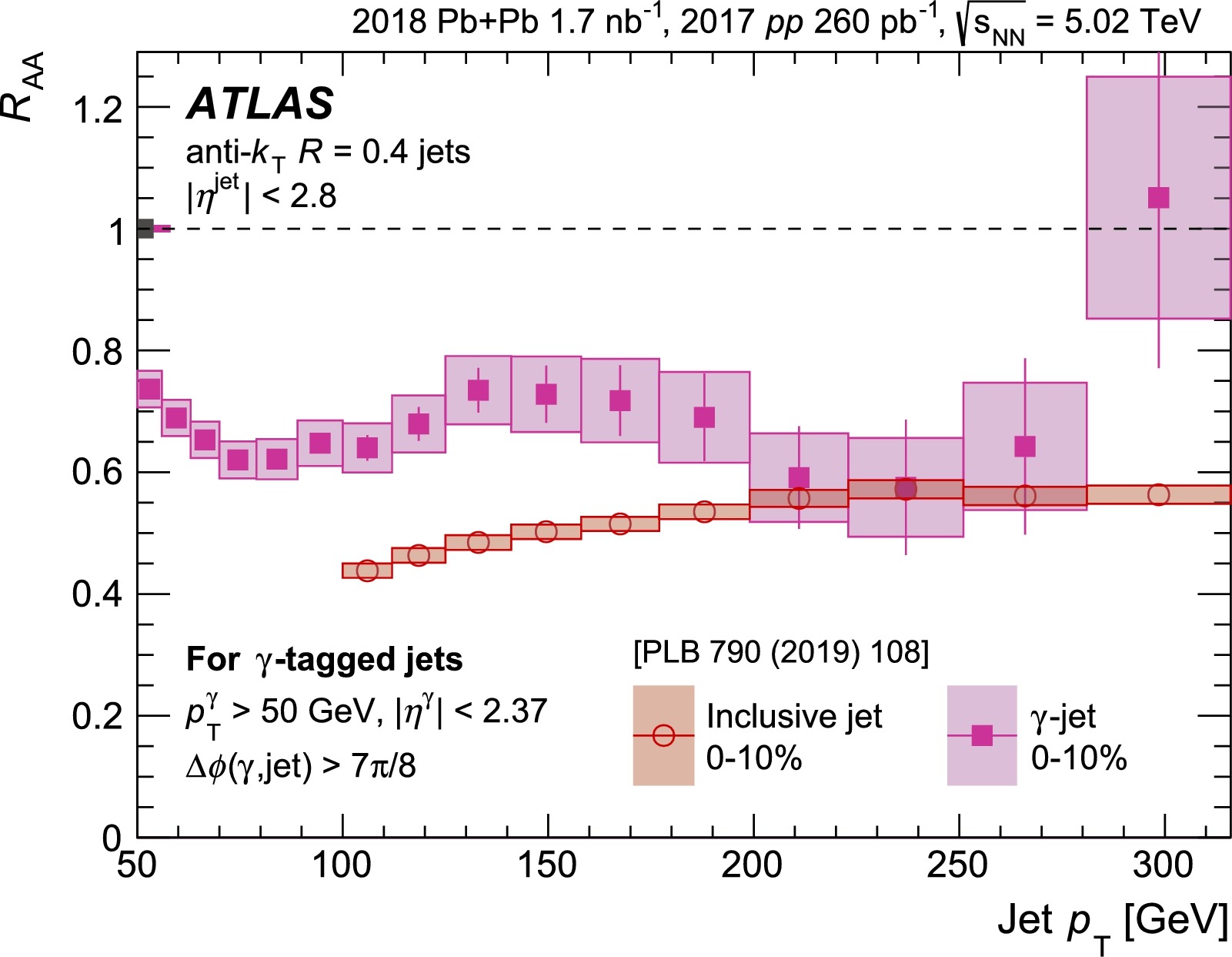}
		\caption{\IAA\ of \gammadir+jet and \pizero+jet from the STAR experiment~\cite{STAR:2023pal,STAR:2023ksv}. \RAA of inclusive jet and \gammadir+jet~\cite{ATLAS:2023iad}. }
        \label{fig:ATLASSTARIAAgammapi0Jet}
	
	\end{center}
\end{figure}

\begin{figure}[htb]
	\begin{center}
		 
		\includegraphics[width=0.35\columnwidth]{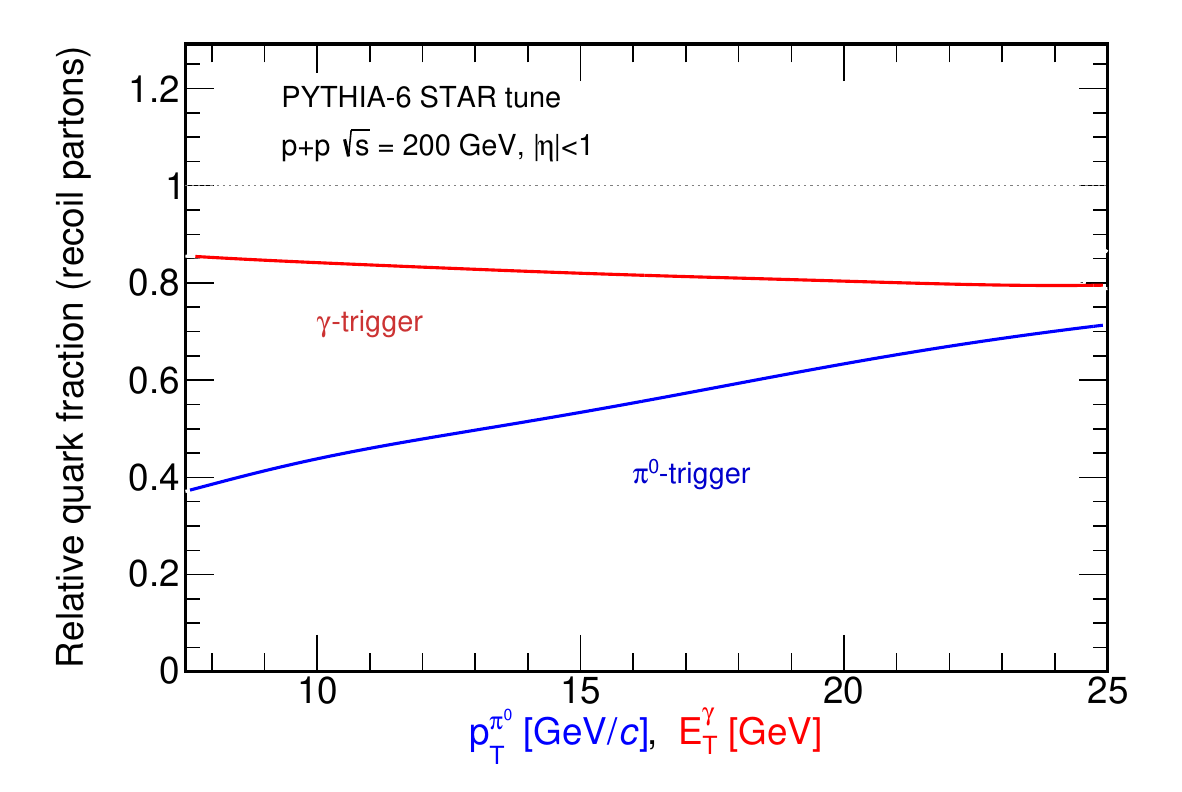}
        \includegraphics[width=0.36\columnwidth]{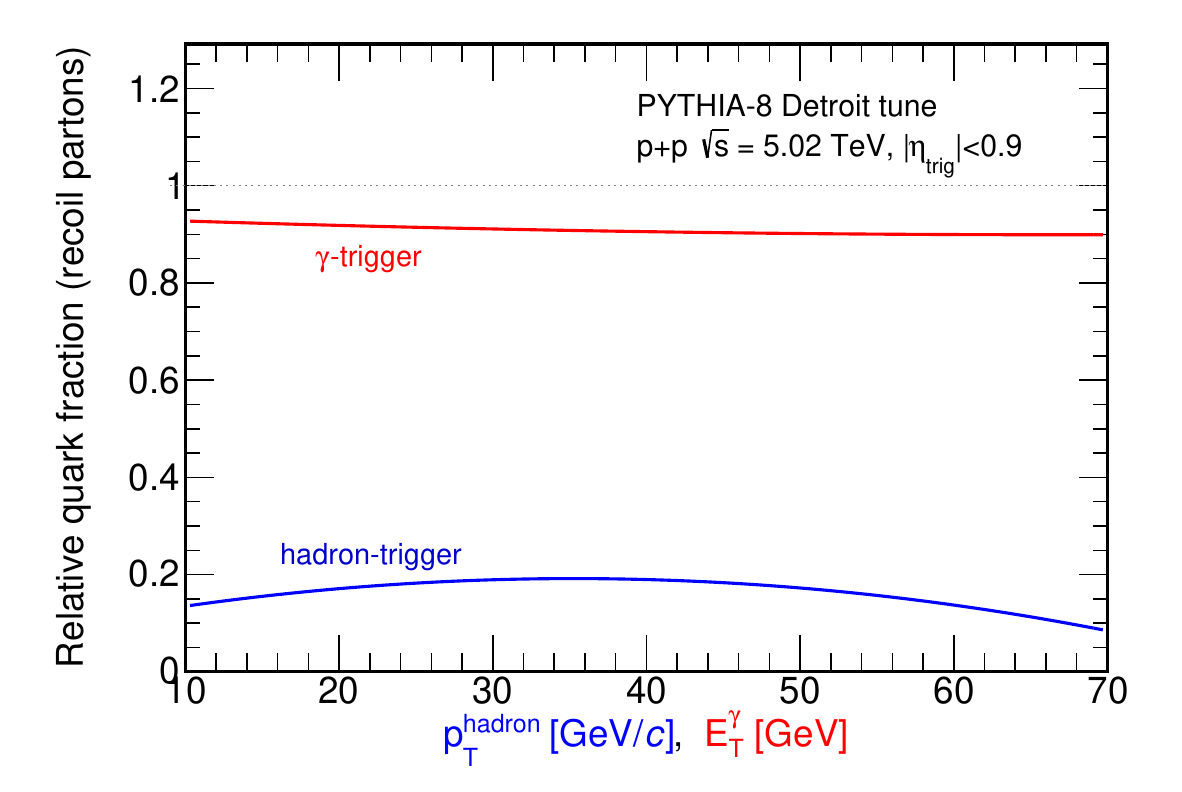}
		\caption{Comparison between quark fraction between RHIC ~\cite{STAR:2023ksv} and LHC energies. } 
        \label{fig:RHICvsLHCqfracComp1}
	\end{center}
\end{figure}

\subsection{Color factor and path length dependence of parton energy loss}\label{Sec::ColorAndPath}

Figure~\ref{fig:ATLASSTARIAAgammapi0Jet} shows the nuclear modification factor (\IAA\ and \RAA) as a function \pTjet, both of which are observed to be less than unity, implying jet yield suppression in central heavy-ion collisions. Jet suppression depends on the path lenght traversed in the medium and flavor of the parton---quark or gluon---due to their different color charges, with gluons ($C_{A}$=3)  losing more energy in the QGP than quarks ($C_{F}$
 =4/3) leading to a larger suppression for gluon-initiated jets. \\
 
 Figures~\ref{fig:RHICvsLHCqfracComp1} show the quark-jet fraction at the LHC and RHIC energies obtained from PYTHIA (with different tunes).
 At LHC energies, a clear difference is observed between the suppression of inclusive jets and 
\gammadir +jets, the latter being predominantly quark-initiated and thus less suppressed. In contrast, STAR measurements at RHIC show, within large uncertainty, no difference between \pizero+jet and \gammadir+jet.
 This could be due to the fact that both trigger jets have comparable quark-jet fraction at these lower energies, different $Q^{2}$, and also lack of precision measurement could not differentiate this difference. \\

 Furthermore, the \RAA\ measurement of b-jets compared to inclusive jets by ATLAS (shown at HP2024) shows that b-jets are approximately 20\% less suppressed than inclusive jets. This supports the expectation that heavy-flavor jets lose less energy in the QGP than light-flavor jets, consistent with the expected mass-dependent energy loss mechanism.\\
 
 Additionally, STAR measurements of \gammadir+jet and \pizero+jet reveal a dependence of jet suppression on $R$ in central Au+Au collisions, suggesting that the lost energy is  recovered at larger $R=0.5$ than the smaller $ R=0.2$ in the medium.\\

\begin{figure}[htb]
	\begin{center}
			 
        \includegraphics[width=0.55\columnwidth]{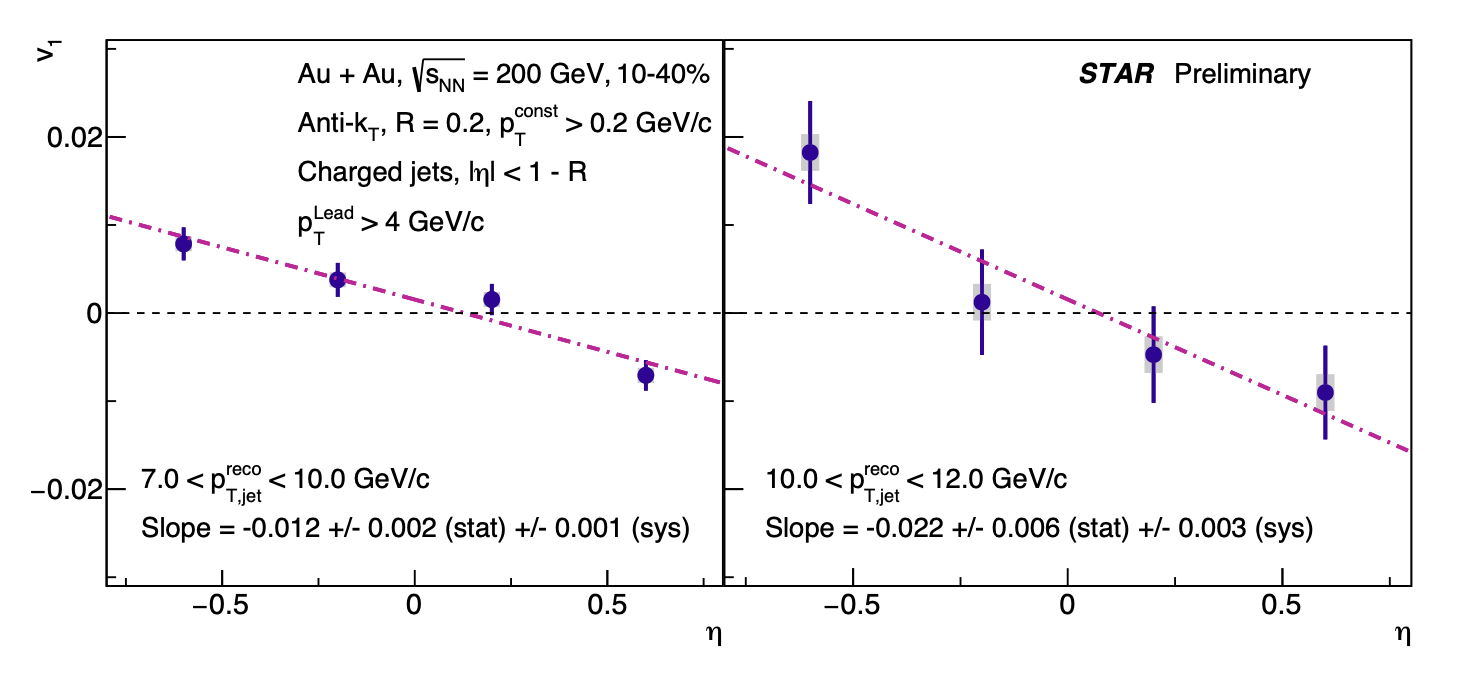}
        \includegraphics[width=0.4\columnwidth]{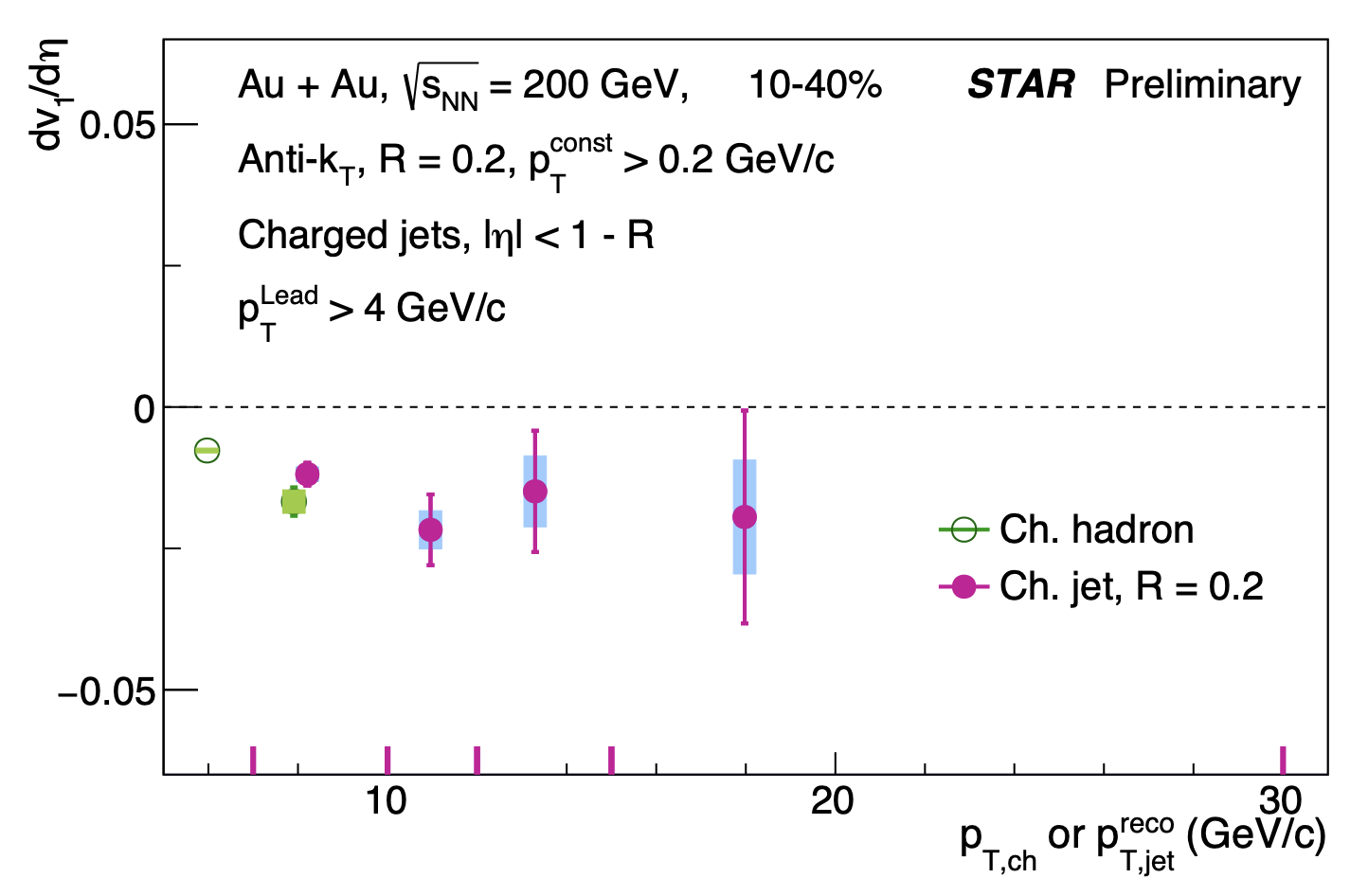}         
        \caption{Jet $v_{1}$ measured by the STAR experiment. Left:  jet $v_{1}$ vs. $\eta$, Right: d$v_{1}$/d$\eta$ vs. \pT.} 
        \label{fig:jetv1}
	\end{center}
\end{figure}

 Figure~\ref{fig:jetv1} shows that the jet $v_{1}$ is an order of magnitude larger than the bulk $v_{1}$, indicating a strong path-length dependence of parton energy loss. This is attributed to the tilted geometry of the fireball produced in non-central Au+Au collisions.
\subsection{Intra-jet broadening due to in-medium radition}
\label{Sec::JetBroadening}

\begin{figure}[htb]
	\begin{center}
		 \includegraphics[width=0.4\columnwidth]{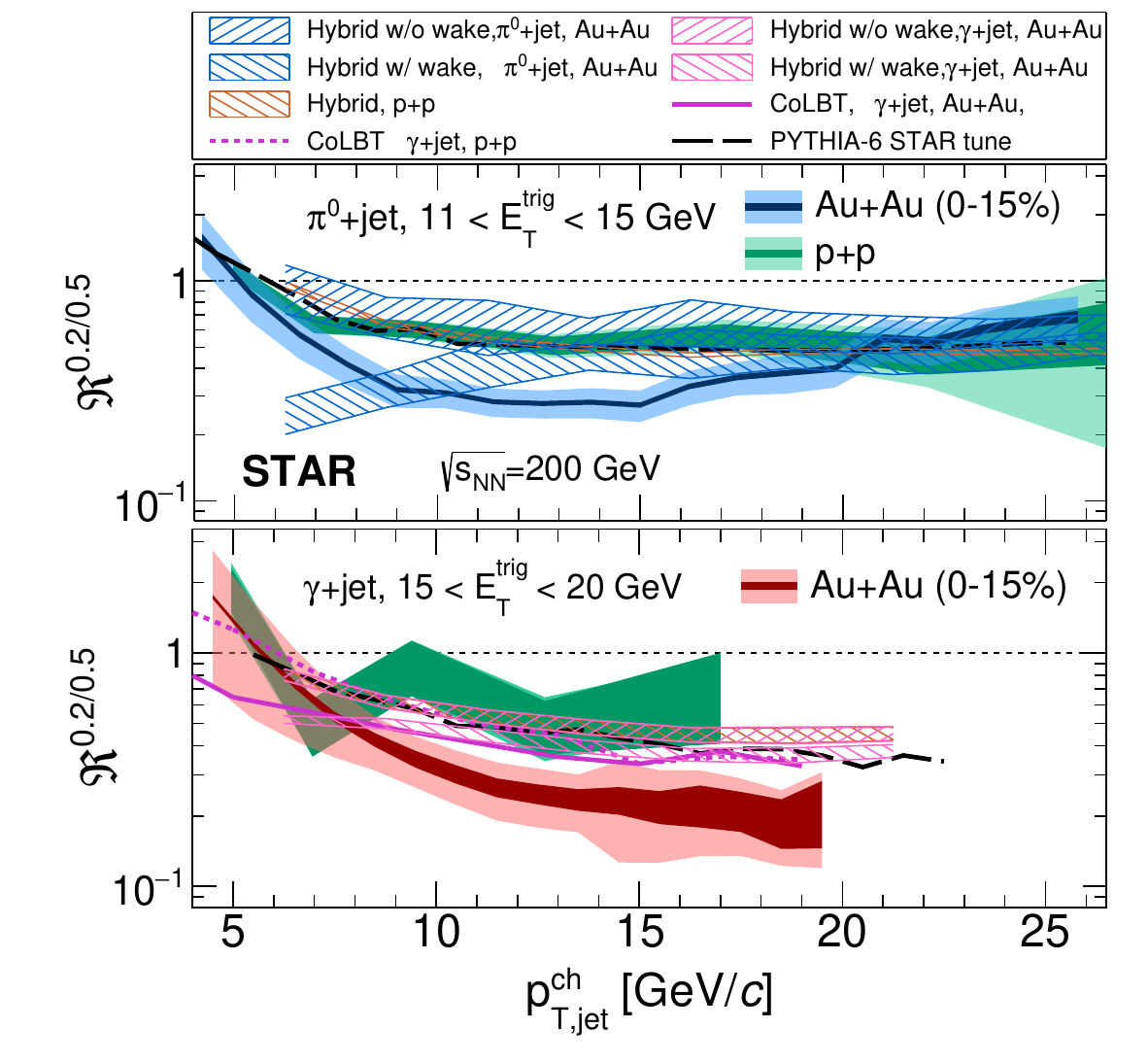}
        \includegraphics[width=0.44\columnwidth]{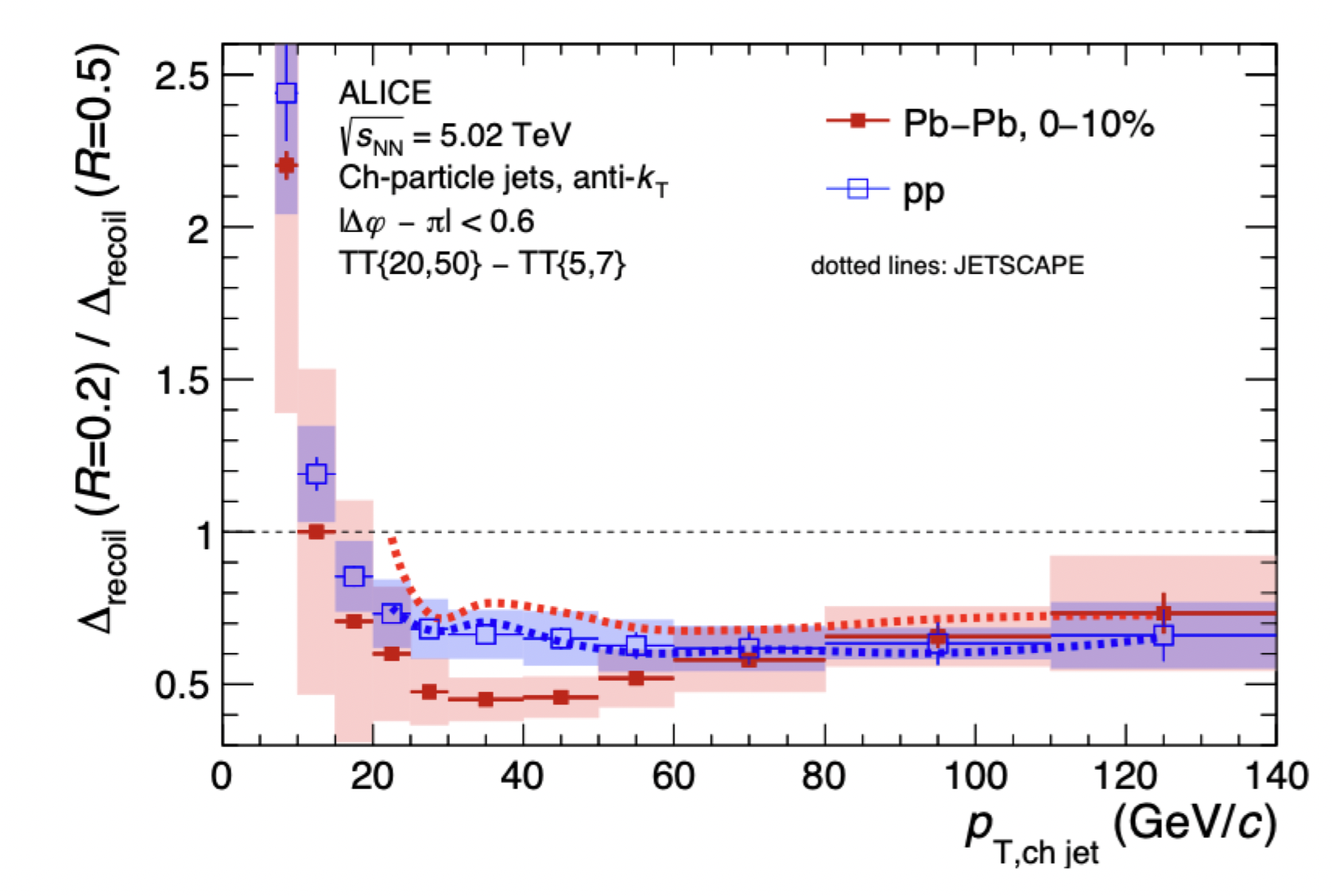}
		\caption{Yield ratios between R=0.2/0.5 for \gammadir+jet and \pizero+jet from the STAR experiment~\cite{STAR:2023pal,STAR:2023ksv}; hadron+jet measurement from ALICE~\cite{ALICE:2023jye,ALICE:2023qve}. } 
        \label{fig:YieldRationppAA}
		
	\end{center}
\end{figure}

Figures~\ref{fig:YieldRationppAA} show the jet yield ratios for R=0.2/0.5 in 
\gammadir +jet and \pizero +jet events from STAR, and h+jet events from ALICE. These ratios, plotted as a function of \pTjet, are found to be systematically lower in heavy-ion collisions than in pp collisions at intermediate \pTjet, suggesting significant intra-jet broadening due to in-medium gluon radiation. This broadening exceeds that expected from vacuum-like radiation in pp, and current theoretical models fail to fully reproduce the observed effect, warranting further theoretical investigation. It is worth to not that the rising trend at high \pTjet\ could be due to the energy loss of trigger particle (high-\pT\ \pizero\ and hadron) in the medium[~\cite{He:2024rcv}], hence such effects is not seen in \gammadir+jet.\\

In addition, the rising trend of yield ratio between $R$=0.2/0.5 at low \pTjet\ region in these semi-inclusive recoil jet measurements differ from the inclusive measurements. This rising tend suggests that at low \pTjet, jets are fatter than at high-\pTjet. A more detailed investigation, including comparison of pA collisions, may provide further insight into origin of this broadening and the QCD evolution. 

\begin{figure}[htb]
	\begin{center}		    \includegraphics[width=0.4\columnwidth]{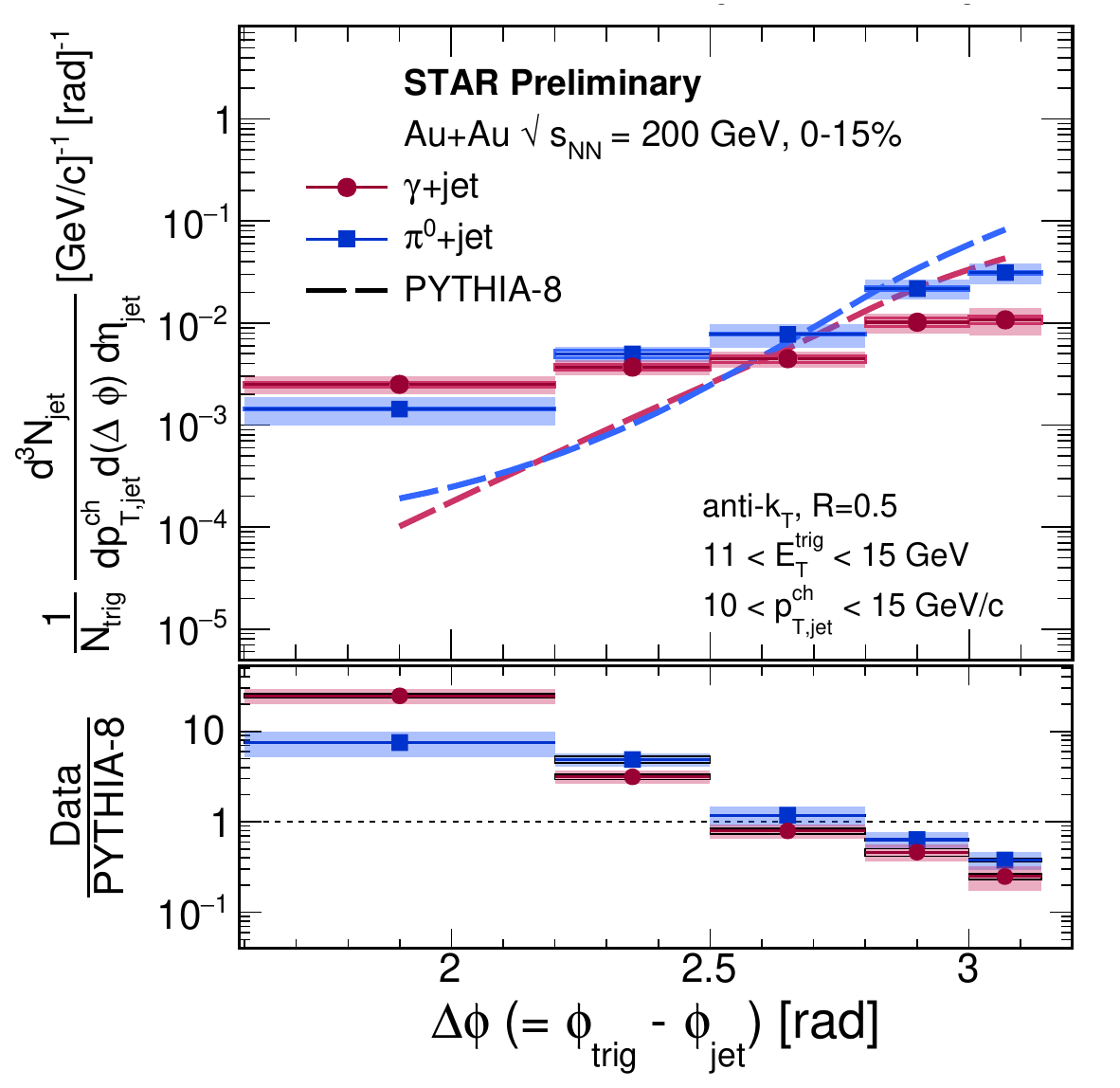}    
	    \includegraphics[width=0.55\columnwidth]{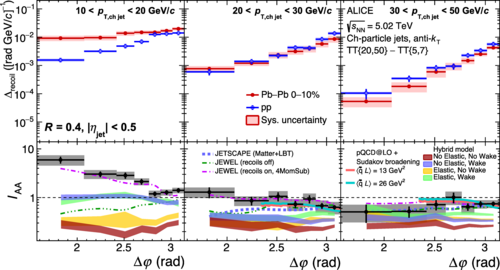}
        \caption{\IAA\ as a function $\Delta\phi$ of \gammadir+jet and \pizero+jet from the STAR~\cite{Sahoo:2023gho} and ALICE ~\cite{ALICE:2023jye,ALICE:2023qve} experiment. } 
        \label{fig:DelphiYieldRationppAA}	
	\end{center}
\end{figure}

\subsection{Medium induced jet acoplanarity in heavy-ion collisions}\label{Sec::JetAcoplanarity}

Due to interactions with the medium, the jet axis can become smeared relative to the direction of its original parent parton, depending on the properties of the QGP. This phenomenon is referred to as jet acoplanarity in heavy-ion collisions. 
Jet acoplanarity is predicted to arise from  Rutherford-like scattering off quasi-particles in the QGP~\cite{DEramo:2012uzl},  the formation of diffusion wake in the medium and/or relatively large in-medium gluon radiations compared to vacuum~\cite{Mueller:2016gko}.\\

Evidence for jet acoplanarity has been observed by STAR through \gammadir+jet and \pizero+jet measurements and h+jet measurement by the ALICE experiment. Figures~\ref{fig:DelphiYieldRationppAA} show the recoil charged jet yields ratio as a function of $\Delta\phi$ ($=\phi^{\rm recoil~jet} - \phi^{\rm trig}$), \IAA ($\Delta\phi$), for different trigger particles from RHIC to the LHC energies. Jets reconstructed with larger resolution parameters $R$ exhibit a strong enhancement at large angles in contrast to jets with smaller $R$ showing yield suppression at all $\Delta\phi$ bins. This enhancement shows a strong dependence on the recoil jet (\pTjet). These observations need further investigation to understand the jet acoplanarity in the medium.


\section{Summary and outlook}\label{Sec::summary}

Recent jet measurements in heavy-ion collisions have provided deeper insights of the distribution of lost parton energy in the QGP, revealing its dependence on color charge (with gluons losing more energy than quarks, path length through the medium, and parton flavor (light vs. heavy). Observations of jet acoplanarity in both STAR and ALICE suggest contributions from medium response effects, underscoring the need for further experimental investigation. A comprehensive understanding of cold nuclear matter effects and baseline vacuum radiation also remains important for interpreting heavy-ion observables. While promising techniques such as Bayesian inference and machine learning are being explored to extract medium transport properties, they are beyond the scope of this proceedings. Similarly, jet substructure measurements and their medium modifications—which offer sensitive probes of the microscopic dynamics of jet–medium interactions—are not addressed in this proceedings. Continued exploration using diverse experimental approaches may not only refine our understanding of QCD matter but also uncover novel phenomena in strongly interacting systems. \\

From 2023 to 2025, the Relativistic Heavy Ion Collider (RHIC) is in a key phase of data taking with its STAR and sPHENIX experiments, both generating high-quality datasets crucial for advancing jet physics in heavy-ion collisions. STAR is conducting detailed measurements of jet quenching and medium response, while also exploring jet substructure using full azimuthal jet reconstruction and correlation techniques. The newly commissioned sPHENIX detector, with its high-granularity electromagnetic and hadronic calorimeters and enhanced tracking, is optimized for precision measurements of jets and photon-tagged jets, enabling detailed studies of jet substructure observables such as angularities, groomed jet mass, and the splitting kinematics of parton showers. Concurrently, the LHC Run 3 is delivering a wealth of complementary data from ALICE, ATLAS, and CMS at significantly higher center-of-mass energies, providing insights into jet substructure modifications and color coherence effects in a hotter and denser QCD medium. Together, the RHIC and LHC programs offer a synergistic and energy-dependent view of partonic energy loss mechanisms and the microscopic structure of the quark-gluon plasma.





\bibliographystyle{cas-model2-names}

\bibliography{cas-refs}

\begin{thebibliography}{18}
\expandafter\ifx\csname natexlab\endcsname\relax\def\natexlab#1{#1}\fi
\providecommand{\url}[1]{\texttt{#1}}
\providecommand{\href}[2]{#2}
\providecommand{\path}[1]{#1}
\providecommand{\DOIprefix}{doi:}
\providecommand{\ArXivprefix}{arXiv:}
\providecommand{\URLprefix}{URL: }
\providecommand{\Pubmedprefix}{pmid:}
\providecommand{\doi}[1]{\href{http://dx.doi.org/#1}{\path{#1}}}
\providecommand{\Pubmed}[1]{\href{pmid:#1}{\path{#1}}}
\providecommand{\bibinfo}[2]{#2}
\ifx\xfnm\relax \def\xfnm[#1]{\unskip,\space#1}\fi
\bibitem[{Aad et~al.(2023)}]{ATLAS:2023iad}
\bibinfo{author}{Aad, G.}, et~al. (\bibinfo{collaboration}{ATLAS}),
  \bibinfo{year}{2023}.
\newblock \bibinfo{title}{{Comparison of inclusive and photon-tagged jet
  suppression in 5.02 TeV Pb+Pb collisions with ATLAS}}.
\newblock \bibinfo{journal}{Phys. Lett. B} \bibinfo{volume}{846},
  \bibinfo{pages}{138154}.
\newblock \DOIprefix\doi{10.1016/j.physletb.2023.138154},
  \href{http://arxiv.org/abs/2303.10090}{\tt arXiv:2303.10090}.
\bibitem[{Aboona et~al.(2023)}]{STAR:2023pal}
\bibinfo{author}{Aboona, B.E.}, et~al. (\bibinfo{collaboration}{STAR}),
  \bibinfo{year}{2023}.
\newblock \bibinfo{title}{{Measurement of in-medium jet modification using
  direct photon+jet and $\pi^{0}$+jet correlations in $p+p$ and central Au+Au
  collisions at $\sqrt{s_{\rm NN}} = 200$ GeV}}
  \href{http://arxiv.org/abs/2309.00156}{\tt arXiv:2309.00156}.
\bibitem[{Acharya et~al.(2020)}]{ALICE:2019qyj}
\bibinfo{author}{Acharya, S.}, et~al. (\bibinfo{collaboration}{ALICE}),
  \bibinfo{year}{2020}.
\newblock \bibinfo{title}{{Measurements of inclusive jet spectra in pp and
  central Pb-Pb collisions at $\sqrt{s_{\rm{NN}}}$ = 5.02 TeV}}.
\newblock \bibinfo{journal}{Phys. Rev. C} \bibinfo{volume}{101},
  \bibinfo{pages}{034911}.
\newblock \DOIprefix\doi{10.1103/PhysRevC.101.034911},
  \href{http://arxiv.org/abs/1909.09718}{\tt arXiv:1909.09718}.
\bibitem[{Acharya et~al.(2022)}]{ALargeIonColliderExperiment:2021mqf}
\bibinfo{author}{Acharya, S.}, et~al. (\bibinfo{collaboration}{A Large Ion
  Collider Experiment, ALICE}), \bibinfo{year}{2022}.
\newblock \bibinfo{title}{{Measurement of the groomed jet radius and momentum
  splitting fraction in pp and Pb$-$Pb collisions at $\sqrt{s_{NN}} = 5.02$
  TeV}}.
\newblock \bibinfo{journal}{Phys. Rev. Lett.} \bibinfo{volume}{128},
  \bibinfo{pages}{102001}.
\newblock \DOIprefix\doi{10.1103/PhysRevLett.128.102001},
  \href{http://arxiv.org/abs/2107.12984}{\tt arXiv:2107.12984}.
\bibitem[{Acharya et~al.(2024a)}]{ALICE:2023jye}
\bibinfo{author}{Acharya, S.}, et~al. (\bibinfo{collaboration}{ALICE}),
  \bibinfo{year}{2024}a.
\newblock \bibinfo{title}{{Measurements of jet quenching using semi-inclusive
  hadron+jet distributions in pp and central Pb-Pb collisions at sNN=5.02
  TeV}}.
\newblock \bibinfo{journal}{Phys. Rev. C} \bibinfo{volume}{110},
  \bibinfo{pages}{014906}.
\newblock \DOIprefix\doi{10.1103/PhysRevC.110.014906},
  \href{http://arxiv.org/abs/2308.16128}{\tt arXiv:2308.16128}.
\bibitem[{Acharya et~al.(2024b)}]{ALICE:2023qve}
\bibinfo{author}{Acharya, S.}, et~al. (\bibinfo{collaboration}{ALICE}),
  \bibinfo{year}{2024}b.
\newblock \bibinfo{title}{{Observation of Medium-Induced Yield Enhancement and
  Acoplanarity Broadening of Low-pT Jets from Measurements in pp and Central
  Pb-Pb Collisions at sNN=5.02\,\,TeV}}.
\newblock \bibinfo{journal}{Phys. Rev. Lett.} \bibinfo{volume}{133},
  \bibinfo{pages}{022301}.
\newblock \DOIprefix\doi{10.1103/PhysRevLett.133.022301},
  \href{http://arxiv.org/abs/2308.16131}{\tt arXiv:2308.16131}.
\bibitem[{Adam et~al.(2015)}]{ALICE:2015mdb}
\bibinfo{author}{Adam, J.}, et~al. (\bibinfo{collaboration}{ALICE}),
  \bibinfo{year}{2015}.
\newblock \bibinfo{title}{{Measurement of jet quenching with semi-inclusive
  hadron-jet distributions in central Pb-Pb collisions at $
  \sqrt{s_{\mathrm{NN}}}=2.76 $ TeV}}.
\newblock \bibinfo{journal}{JHEP} \bibinfo{volume}{09}, \bibinfo{pages}{170}.
\newblock \DOIprefix\doi{10.1007/JHEP09(2015)170},
  \href{http://arxiv.org/abs/1506.03984}{\tt arXiv:1506.03984}.
\bibitem[{Adamczyk et~al.(2016)}]{STAR:2016jdz}
\bibinfo{author}{Adamczyk, L.}, et~al. (\bibinfo{collaboration}{STAR}),
  \bibinfo{year}{2016}.
\newblock \bibinfo{title}{{Jet-like Correlations with Direct-Photon and
  Neutral-Pion Triggers at $\sqrt{s_{_{NN}}} = 200$ GeV}}.
\newblock \bibinfo{journal}{Phys. Lett. B} \bibinfo{volume}{760},
  \bibinfo{pages}{689--696}.
\newblock \DOIprefix\doi{10.1016/j.physletb.2016.07.046},
  \href{http://arxiv.org/abs/1604.01117}{\tt arXiv:1604.01117}.
\bibitem[{Adamczyk et~al.(2017)}]{STAR:2017hhs}
\bibinfo{author}{Adamczyk, L.}, et~al. (\bibinfo{collaboration}{STAR}),
  \bibinfo{year}{2017}.
\newblock \bibinfo{title}{{Measurements of jet quenching with semi-inclusive
  hadron+jet distributions in Au+Au collisions at $\sqrt{s_{NN}}$ = 200 GeV}}.
\newblock \bibinfo{journal}{Phys. Rev. C} \bibinfo{volume}{96},
  \bibinfo{pages}{024905}.
\newblock \DOIprefix\doi{10.1103/PhysRevC.96.024905},
  \href{http://arxiv.org/abs/1702.01108}{\tt arXiv:1702.01108}.
\bibitem[{Adams et~al.(2003)}]{STAR:2003pjh}
\bibinfo{author}{Adams, J.}, et~al. (\bibinfo{collaboration}{STAR}),
  \bibinfo{year}{2003}.
\newblock \bibinfo{title}{{Evidence from d + Au measurements for final state
  suppression of high p(T) hadrons in Au+Au collisions at RHIC}}.
\newblock \bibinfo{journal}{Phys. Rev. Lett.} \bibinfo{volume}{91},
  \bibinfo{pages}{072304}.
\newblock \DOIprefix\doi{10.1103/PhysRevLett.91.072304},
  \href{http://arxiv.org/abs/nucl-ex/0306024}{\tt arXiv:nucl-ex/0306024}.
\bibitem[{Adler et~al.(2007)}]{PHENIX:2006avo}
\bibinfo{author}{Adler, S.S.}, et~al. (\bibinfo{collaboration}{PHENIX}),
  \bibinfo{year}{2007}.
\newblock \bibinfo{title}{{High transverse momentum $\eta$ meson production in
  $p^+ p$, $d^+$ Au and Au+Au collisions at $S(NN) ^{(1/2)}$ = 200-GeV}}.
\newblock \bibinfo{journal}{Phys. Rev. C} \bibinfo{volume}{75},
  \bibinfo{pages}{024909}.
\newblock \DOIprefix\doi{10.1103/PhysRevC.75.024909},
  \href{http://arxiv.org/abs/nucl-ex/0611006}{\tt arXiv:nucl-ex/0611006}.
\bibitem[{Cacciari et~al.(2008)Cacciari, Salam and Soyez}]{Cacciari:2008gp}
\bibinfo{author}{Cacciari, M.}, \bibinfo{author}{Salam, G.P.},
  \bibinfo{author}{Soyez, G.}, \bibinfo{year}{2008}.
\newblock \bibinfo{title}{{The anti-$k_t$ jet clustering algorithm}}.
\newblock \bibinfo{journal}{JHEP} \bibinfo{volume}{04}, \bibinfo{pages}{063}.
\newblock \DOIprefix\doi{10.1088/1126-6708/2008/04/063},
  \href{http://arxiv.org/abs/0802.1189}{\tt arXiv:0802.1189}.
\bibitem[{Collab(2023)}]{STAR:2023ksv}
\bibinfo{author}{Collab, S.} (\bibinfo{collaboration}{STAR}),
  \bibinfo{year}{2023}.
\newblock \bibinfo{title}{{Semi-inclusive direct photon+jet and $\pi^{0}$+jet
  correlations measured in $p+p$ and central Au+Au collisions at
  $\sqrt{s_\mathrm{NN}} = 200$ GeV}} \href{http://arxiv.org/abs/2309.00145}{\tt
  arXiv:2309.00145}.
\bibitem[{D'Eramo et~al.(2013)D'Eramo, Lekaveckas, Liu and
  Rajagopal}]{DEramo:2012uzl}
\bibinfo{author}{D'Eramo, F.}, \bibinfo{author}{Lekaveckas, M.},
  \bibinfo{author}{Liu, H.}, \bibinfo{author}{Rajagopal, K.},
  \bibinfo{year}{2013}.
\newblock \bibinfo{title}{{Momentum Broadening in Weakly Coupled Quark-Gluon
  Plasma (with a view to finding the quasiparticles within liquid quark-gluon
  plasma)}}.
\newblock \bibinfo{journal}{JHEP} \bibinfo{volume}{05}, \bibinfo{pages}{031}.
\newblock \DOIprefix\doi{10.1007/JHEP05(2013)031},
  \href{http://arxiv.org/abs/1211.1922}{\tt arXiv:1211.1922}.
\bibitem[{Dokshitzer et~al.(1997)Dokshitzer, Leder, Moretti and
  Webber}]{Dokshitzer:1997in}
\bibinfo{author}{Dokshitzer, Y.L.}, \bibinfo{author}{Leder, G.D.},
  \bibinfo{author}{Moretti, S.}, \bibinfo{author}{Webber, B.R.},
  \bibinfo{year}{1997}.
\newblock \bibinfo{title}{{Better jet clustering algorithms}}.
\newblock \bibinfo{journal}{JHEP} \bibinfo{volume}{08}, \bibinfo{pages}{001}.
\newblock \DOIprefix\doi{10.1088/1126-6708/1997/08/001},
  \href{http://arxiv.org/abs/hep-ph/9707323}{\tt arXiv:hep-ph/9707323}.
\bibitem[{He et~al.(2024)He, Nie, Cao, Ma, Yi and Caines}]{He:2024rcv}
\bibinfo{author}{He, Y.}, \bibinfo{author}{Nie, M.}, \bibinfo{author}{Cao, S.},
  \bibinfo{author}{Ma, R.}, \bibinfo{author}{Yi, L.}, \bibinfo{author}{Caines,
  H.}, \bibinfo{year}{2024}.
\newblock \bibinfo{title}{{Deciphering yield modification of hadron-triggered
  semi-inclusive recoil jets in heavy-ion collisions}}.
\newblock \bibinfo{journal}{Phys. Lett. B} \bibinfo{volume}{854},
  \bibinfo{pages}{138739}.
\newblock \DOIprefix\doi{10.1016/j.physletb.2024.138739},
  \href{http://arxiv.org/abs/2401.05238}{\tt arXiv:2401.05238}.
\bibitem[{Mueller et~al.(2016)Mueller, Wu, Xiao and Yuan}]{Mueller:2016gko}
\bibinfo{author}{Mueller, A.H.}, \bibinfo{author}{Wu, B.},
  \bibinfo{author}{Xiao, B.W.}, \bibinfo{author}{Yuan, F.},
  \bibinfo{year}{2016}.
\newblock \bibinfo{title}{{Probing Transverse Momentum Broadening in Heavy Ion
  Collisions}}.
\newblock \bibinfo{journal}{Phys. Lett. B} \bibinfo{volume}{763},
  \bibinfo{pages}{208--212}.
\newblock \DOIprefix\doi{10.1016/j.physletb.2016.10.037},
  \href{http://arxiv.org/abs/1604.04250}{\tt arXiv:1604.04250}.
\bibitem[{Sahoo(2024)}]{Sahoo:2023gho}
\bibinfo{author}{Sahoo, N.R.} (\bibinfo{collaboration}{STAR}),
  \bibinfo{year}{2024}.
\newblock \bibinfo{title}{{STAR Experimental Highlights at Hard Probes 2023}}.
\newblock \bibinfo{journal}{PoS} \bibinfo{volume}{HardProbes2023},
  \bibinfo{pages}{006}.
\newblock \DOIprefix\doi{10.22323/1.438.0006},
  \href{http://arxiv.org/abs/2308.04801}{\tt arXiv:2308.04801}.

\end{thebibliography}



\end{document}